\documentclass[aps,prb,10pt,notitlepage,twocolumn]{revtex4-1}
\usepackage{amsmath}
\usepackage{amssymb}
\usepackage{amsthm}
\usepackage{amsfonts}
\usepackage{bbm}
\usepackage{tensor}
\usepackage{dsfont}
\usepackage{graphicx}
\usepackage[section]{placeins}
\usepackage{mathrsfs}
\usepackage{stmaryrd}
\usepackage[all]{xy}
\usepackage[mathcal]{eucal}
\usepackage{verbatim}  
\usepackage[cm]{fullpage}  
\usepackage{wasysym}
\usepackage{url}
\usepackage{csquotes}
\MakeOuterQuote{"}
\usepackage{subcaption}
\usepackage{ulem}
\usepackage{enumerate}
\newcommand{\bA}{\mathbf{A}}
\newcommand{\bB}{\mathbf{B}}

\newcommand{\bP}{\mathbf{P}}

\newcommand{\bS}{\mathbf{S}}
\newcommand{\bT}{\mathbf{T}}

\newcommand{\bk}{\mathbf{k}}
\newcommand{\bq}{\mathbf{q}}

\newcommand{\bzero}{\mathbf{0}}

\newcommand{\bpsi}{\boldsymbol\psi}

\newcommand{\be}{{\bf e}}
\newcommand{\bom}{{\bf m}}

\newcommand{\fD}{\mathcal{D}}

\newcommand{\dd}{{\rm d}}
\newcommand{\abs}[1]{\left| #1 \right|}
\newcommand{\av}[1]{\left\langle #1 \right\rangle}
\newcommand{\ket}[1]{\left| #1 \right\rangle}

\newcommand{\inprod}[2]{\left\langle #1 \middle | #2 \right\rangle}
\newcommand{\inoprod}[3]{\left\langle #1 \middle | #2 \middle| #3 \right\rangle}

\newcommand{\aS}{\mathcal{S}}

\begin{document}
\title{Theory of excitations and dielectric response at a spin-orbital quantum critical point}
\author{Daniel Ish}
\affiliation{Physics Department, University of California, Santa Barbara, California 93106, USA}
\author{Leon Balents}
\affiliation{Kavli Institute for Theoretical Physics, University of California, Santa Barbara, California 93106, USA}
\begin{abstract}
Despite possessing a local spin $2$ moment on the iron site and a Curie-Weiss temperature of $45K$, the A site spinel FeSc$_2$S$_4$ does not magnetically order down to 50mK.\cite{CW,exp-review} Previous theoretical work\cite{fescs-prb,fescs-prl} by Chen and Balents advanced an explanation for this observation in the form of the ``$J_2$-$\lambda$'' model which places FeSc$_2$S$_4$ close to a quantum critical point on the disordered side of a quantum phase transition between a N\'{e}el ordered phase and a "Spin-Orbital Liquid" in which spins and orbitals are entangled, quenching the magnetization. We present new theoretical studies of the optical properties of the $J_2$-$\lambda$ model, including a computation of the dispersion relation for the quasiparticle excitations and the form of the collective response to electric field.  We argue that the latter directly probes a low energy excitation continuum characteristic of quantum criticality, and that our results reinforce the consistency of this model with experiment.
\end{abstract}
\maketitle
\section{Introduction}

A suite of experimental probes\cite{ex-cont,ex-gfac,kalvius2006musr,buttgen2006spin,neutrons,CW,exp-review} identifies the A-site spinel FeSc$_2$S$_4$ as a rare example of a orbitally degenerate antiferromagnet which resists magnetic or orbital order down to a temperature of tens of millikelvin, making it a truly quantum paramagnet.  It has been suggested\cite{fescs-prb,fescs-prl} to lie close to a quantum critical point, making it a compelling object of study. This prior theoretical work proposed the "$J_2$-$\lambda$" model for this compound in terms of spin two, $\bS_j$, and spin one half, $\bT_j$, operators on the diamond lattice with the Hamiltonian
\begin{align}
H = J_2\sum_{\langle i,j\rangle}\bS_i\cdot\bS_j + \sum_{i}\mathcal{H}^0_i - B\sum_iS^z_i\label{eq:1}
\end{align}
where $\langle i ,j\rangle$ stands for next nearest neighbor bonds and the on-site Hamiltonian $\mathcal{H}^0_i$ is given by
\begin{align}
\mathcal{H}^0_i = -\frac{\lambda}{3}\left(\sqrt{3}T^x_i\left[\left(S^x_i\right)^2 -\left(S^y_i\right)^2 \right] + T^z_i\left[3\left(S^z_i\right)^2 - \bS_i^2\right]\right)
\end{align}
with $J_2>0$ and $\lambda > 0$. We have included in Eq.~\eqref{eq:1} an external magnetic field $B$, taken for concreteness along the crystalline $(001)$ axis, whose effects we will study further in the following.  One should interpret this Hamiltonian as describing the low-energy dynamics of the $6$ $d$ electrons on the Fe$^{2+}$ sites. Due to the tetrahedral crystal field, the $d$ manifold splits into a lower $e$ doublet and an upper $t_{2}$ triplet. These states are then filled in a high-spin configuration on the assumption that on site Hund's Rule exchange dominates the crystal field splitting, giving an overall spin 2 ($\bS_i$) together with a two fold orbital degeneracy ($\bT_i$). The $J_2$ term in the Hamiltonian is a NNN antiferromagnetic exchange term which arises in the standard way from virtual hopping between Fe$^{2+}$ sites and the $\lambda$ term represents the effect of spin-orbit coupling at second order, coupling the $e$ hole to the overall spin $2$.\cite{opticalabFe-PR,opticalabFe-PRB}

With $J_2 = B = 0$, $\mathcal{H}^0_i$ describes a system of uncoupled Fe$^{2+}$ sites with tetrahedral geometry under the influence of spin orbit coupling. This splits the 10-fold degenerate high-spin manifold into five equally spaced levels separated by energy $\lambda$. These are, in order of ascending energy, an $a_1$ singlet, a $t_1$ triplet, an $e$ doublet, a $t_2$ triplet and an $a_2$ singlet. The $a_1$ ground state takes the form
\begin{align}
\frac{1}{\sqrt{2}}\ket{x^2 - y^2}\ket{0} + \frac{1}{2}\ket{3z^2 - r^2}\left(\ket{2}+\ket{-2}\right)
\end{align}
where the number in the second ket in each product refers to the $S^z$ eigenvalue. Critically, this state has zero average magnetization along all axes. In the presence of $J_2$, single site $t_1$ excitations ("triplons" or "spin-orbitons") acquire a $\bk$-dependent dispersion, but remain massive until $J_2/\lambda = 1/16$ at which point the system undergoes a quantum phase transition to antiferromagnetic ordering at wave vector $\bq = (2\pi,0,0)$ and symmetry related wave vectors.\cite{fescs-prb,fescs-prl} In the $J_2$-$\lambda$ model, this ordering actually happens independently on each of the fcc sublattices. A NN exchange term, $J_1$, is also allowed by symmetry,\cite{fescs-prb,fescs-prl} but is expected to be much smaller\cite{fescs-prb,fescs-prl,DFT} and is difficult to distinguish from $J_2$ experimentally at $\bk = 0$, which will be the regime of focus in this article. This term controls the relative orientation of the magnetizations of the $A$ and $B$ sublattices.

\begin{figure}[hbtp]
\includegraphics[scale=1.2]{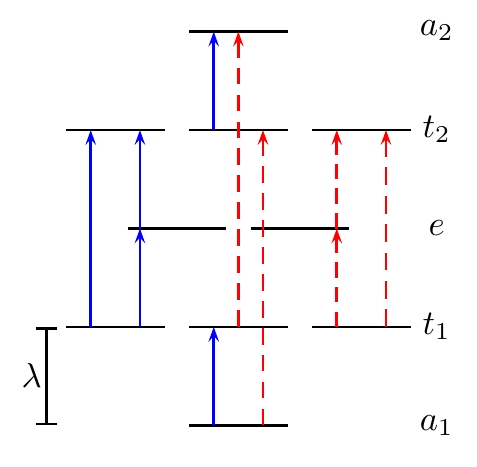}
\caption{Spectrum of $\mathcal{H}^0_i$, with magentic dipole allowed transitions (blue) and electric dipole allowed transitions (red, dashed). The number of lines in a level indicates the degeneracy of that level.}
\end{figure}

One can argue for the consistency of this model with FeSc$_2$S$_4$ at lowest order in terms of two distinctive experimental observations: the lack of observed magnetic ordering down to temperatures of $50$mK\cite{CW,exp-review} and the observation of a low energy mode at momentum $(2\pi,0,0)$ by neutron scattering.\cite{neutrons} Taken together, these suggest that the model at least qualitatively predicts the behavior of the compound if we suppose that the compound lies on the disordered side of the transition close to the QCP. Indeed, previous estimates\cite{fescs-prb,fescs-prl} of the magnitude of these couplings in FeSc$_2$S$_4$ give $\lambda\approx 22.1$ K and $J_2\approx 1.29$ K, giving $J_2/\lambda\approx 1/17$ and putting the material just on the disordered side of the transition.

Given the striking properties of this material, it is no surprise that it has been the subject of recent experimental investigations, primarily focused on its optical properties.\cite{ex-cont,ex-gfac} In this article, we attempt to make  contact between this model and the observed optical properties of FeSc$_2$S$_4$ in order to argue for its continued consistency with experimental results. In order to do this, we will both investigate the fate of single site dipole allowed transitions in the presence of non-zero $J_2$, predicting both the shift in location and $g$ factor, and investigate the character of the collective response to $\bk = 0$ electric fields. We also derive expressions for as-yet unobserved quantities and propose possible experiments to measure them.

\section{Magnetic Dipole Exitations}
The single site problem has a single magnetic dipole allowed transition from the ground state, $a_1\rightarrow t_1$. We expect that in the presence of $J_2$, this excitation will still result in a peak in the AC response of the material to magnetic fields, but will shift in energy. Given the proposed proximity of the material to a quantum phase transition, we expect perturbation theory to be inaccurate when predicting the location of these excitations. To see this, note that corrections to the energies of low lying excitations due to $J_2$ must be comparable to $\lambda$ close to the QCP, since the $J_2 = 0$ gap (of magnitude $\lambda$) has almost closed. Thus, to determine their dispersion, we instead calculate the RPA susceptibility in the presence of a (001) directed field and investigate its pole structure as a function of $\bk$.

As we will see in the present section, this corresponds to treating the $t_1$ excitations as non-interacting boson excitations (triplons) above a variational ground state (the mean field). On a practical level, this means that we will keep a smattering of selected terms at higher orders in $J_2$. This approximation ultimately relies on the fluctuations about the mean field being small, i.e. on the triplons being dilute. This approach, of course, has its own dangers. For one, the astute reader will notice that this approximation also breaks down as we approach the critical point, where these triplons condense. The hope here is not that the RPA is perfect close to the critical point, simply that it does better than perturbation theory due to the inclusion of these higher order terms. The natural second concern is that our choice of a description in terms of free bosons was incorrect, i.e. that we have chosen an unfortunate auxiliary field. Put another way, we may have decided on a bad prescription for which terms of the perturbation series to include. As some modest check on this, we will compare the RPA results with those of low order perturbation theory and consider more carefully the validity of this description wherever the two disagree.
\subsection{Formalism and $B = 0$ Magnetic Dipole Excitations}
We consider the imaginary time dynamic susceptibility 
\begin{align}
\chi^{\mu\nu}_{ij}(\tau_1 - \tau_2) = \av{T_{\tau}\aS^{\mu}_i(\tau_1)\aS^\nu_j(\tau_2)}
\end{align}
where
\begin{align}
\aS^\mu(\tau) = S^\mu(\tau) - \av{S^\mu}
\end{align}
Performing a Hubbard-Stratonovich transformation to decouple the exchange term, we find the partition function to be given by
\begin{align}
Z = \int \fD[\Phi] e^{-S_{eff}[\Phi]}
\end{align}
with the effective action for the auxiliary field $\Phi$
\begin{align}
S_{eff}[\Phi]=\frac{1}{2}\int\dd\tau\sum_{ij}J^{-1}_{ij}\Phi_i\cdot\Phi_j -\ln W[\Phi]
\label{eq:Seff}
\end{align} 
where $W[\Phi]$ is the partition function for the $J_2 = 0$ problem with a magnetic field of $-i\Phi^\mu\be^\mu$ applied to each site (the "single site" problem). The dynamic susceptibility is then related to the propagator for the auxiliary field, $\Omega(\bk,\omega)$, by
\begin{align}
\chi(\bk,\omega) = J(\bk)^{-1} - J(\bk)^{-2}\Omega(\bk,\omega)
\label{eq:relate}
\end{align}
where $J(\bk)$ is the Fourier transform of the interaction
\begin{align}
J(\bk) =  J_2 \sum_{\bA}\cos\left(\bk\cdot \bA\right)
\end{align}
with $\bA$ the 12 fcc nearest neighbors. If we then expand the action to second order in $\Phi$ about its saddle point ($\Phi = 0$), we find that the bare propagator for the auxiliary field, $\Omega_0$, is given by
\begin{align}
\Omega_0 = [J(\bk)^{-1} + \chi_0(\omega)]^{-1}
\end{align}
with $\chi_0(\omega)$ the dynamic susceptibility of the single site problem and the inverse being, of course, the matrix inverse. Together with Equation~\ref{eq:relate} this gives the RPA susceptibility 
\begin{align}
\chi_{RPA}(\bk,\omega) = \chi_0(\omega)[1 + J(\bk)\chi_0(\omega)]^{-1}
\label{eq:rpa}
\end{align}
which we analytically continue to extract the real-time RPA susceptibility in terms of the real-time single site susceptibility. We can write the real time single site susceptibility in terms of the spectral representation 
\begin{align}
\chi_0^{\mu\nu}(\omega) = \sum_{j \neq 0}\frac{\inoprod{0}{S^\mu}{j}\inoprod{j}{S^\nu}{0}}{E_j - E_0 - \omega} +\frac{\inoprod{0}{S^\nu}{j}\inoprod{j}{S^\mu}{0}}{E_j - E_0 + \omega} 
\end{align}
with $\ket{0}$ denoting the ground state. Using this, we find that both the single site and RPA susceptibilities are multiples of the identity and that the RPA susceptibility exhibits poles at 
\begin{align}
\omega(\bk) =  \lambda\sqrt{1 + \frac{4}{\lambda}J(\bk)}
\end{align}
which predicts a pole in the $\bk = 0$ susceptibility at
\begin{align}
\omega(\bzero) = \lambda\sqrt{1 + 48\frac{J_2}{\lambda}}\approx 1.95 \lambda
\label{eq:old}
\end{align} 
at the predicted value for $J_2$, $J_2 = \lambda/17$, as previously reported.\citep{fescs-prb} Since we have neglected all interactions, robbing the triplons of any decay channels, these poles lie on the real-frequency axis and give rise to $\delta$ function peaks in the imaginary part of the real-time susceptibility. Expanding this dispersion to first order in $J_2$ gives
\begin{align}
\omega(\bk) = \lambda + 2J(\bk) + O(J_2^2)
\end{align}
in agreement with first order perturbation theory.
\subsection{Magnetic Dipole Excitations with $B \neq 0$}
\label{sec:magdip}
The preceding analysis can be repeated in the presence of a magnetic field with little change. The location of the $\Phi$ saddle point simply shifts due to the presence of a magnetic term in the associated single site problem, leading to a different mean field. The new saddle point is of the form
\begin{align}
\Phi_0(\tau) = -i \beta^\mu
\end{align}
with $\beta^\mu$ determined by the mean field consistency equation
\begin{align}
\beta^\mu = 12 J_2 \av{S^\mu}_{0}
\end{align}
where $\av{\cdot}_{0}$ stands for averages taken in the single site problem in the presence of the field $\left(B^\mu - \beta^\mu\right)\be^\mu$, with $B^\mu\be^\mu$ the applied field. In our case, since the applied field is considered only along the $(001)$ direction, we have that only $\beta^z$ is non-zero. The form of the RPA susceptibility is the same as in Equation~\ref{eq:rpa}, save for the fact that the single site susceptibility is now calculated in the mean field
\begin{align}
\chi_0^{\mu\nu}(\tau) = \av{T_\tau \aS^\mu(\tau)\aS^\nu(0)}_{0}
\end{align}

Applying the magnetic field reduces the tetrahedral symmetry of the single site problem, causing some transitions $a_1\rightarrow t_2$ to become allowed. As will be discussed in Section~\ref{sec:Eresp}, the shifts in the energies of the $t_2$ excitations may not be well captured by the the $J_2$-$\lambda$ model and have not yet been experimentally observed. In the analysis that follows, the poles due to these transition exhibit some pathological behavior. In particular, as $B\rightarrow 0$, they return to the single site energy of the $t_2$ excitations, in contrast with the behavior of the $t_1$ excitations. This is also in contrast to the perturbative result, which gives a shift of the $t_2$ energies at second order in $J_2$. Consequently, we will not present the results of the $B\neq 0$ RPA for the two $t_2$ excitations that become allowed.

\begin{figure}[t]
\includegraphics[scale=1]{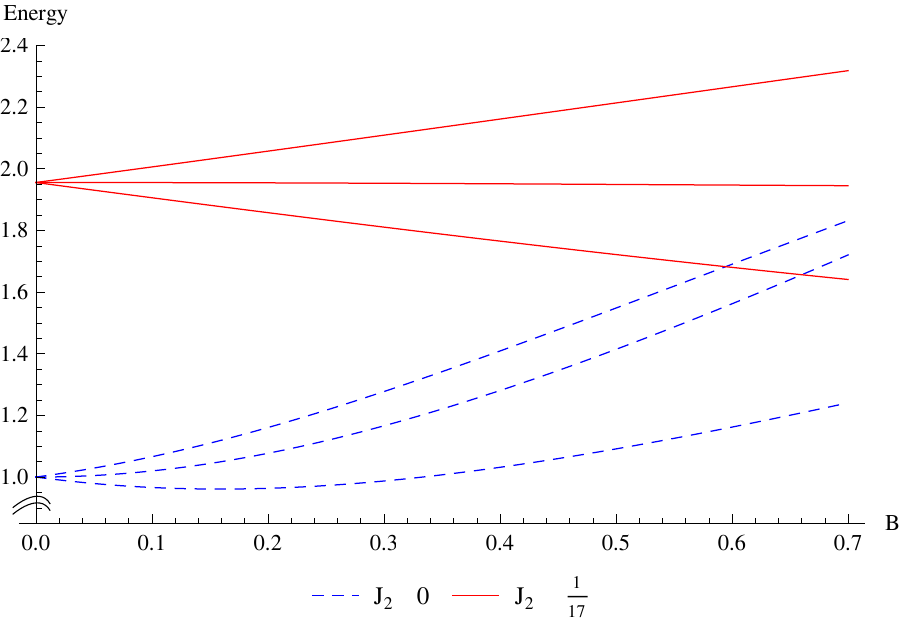}
\caption{Energy of the $t_1$ triplet excitations versus $B$ in the single site problem (dashed) and as given by RPA at $\bk = 0$ (solid). Field and energy are both measured in units of $\lambda$.}
\label{fig:disp}
\end{figure}

For $B\neq 0$, the analysis of the pole structure of the RPA suceptibility is complicated significantly by that fact that the single site susceptibility is no longer a multiple of the identity. We only retain that the susceptibility is block-diagonal in the $zz$ and $xy$ blocks. Even with this reduced symmetry, there is still some remaining structure that we can exploit. Using the spectral representation, we see that $\chi^{xx}_0 = \chi^{yy}_0$ and $\chi^{xy}_0 = -\chi^{yx}_0$. Together with the fact that $\chi^{yx}_0$ is purely imaginary, this implies that the single site susceptibility is diagonalized at all frequencies (and hence all imaginary times) by the same unitary transformation. The eigenvalues of the $xy$ single site susceptibility are then
\begin{align}
\chi_0^{\pm}(\tau) = \chi^{xx}_0(\tau) \pm i\chi^{xy}(\tau)  = \frac{1}{2}\av{T_\tau S^\mp(\tau)S^\pm(0)}_0
\end{align}
and, of course, $\chi^{zz}$. Together with Equation~\ref{eq:rpa}, gives that the eigenvalues of the RPA susceptibility are
\begin{align}
\chi_{RPA}^\pm(\bk,\omega) = \frac{\chi_0^\pm(\omega)}{1 + J(\bk)\chi_0^\pm(\omega)}
\end{align}
and
\begin{align}
\chi_{RPA}^{zz}(\bk,\omega)= \frac{\chi_0^{zz}(\omega)}{1 + J(\bk)\chi_0^{zz}(\omega)}
\end{align}
Furthermore, we can actually see by the spectral representation that $\chi_0^{-}(\omega) = \chi_0^+(-\omega)$, which implies the same result for $\chi^-_{RPA}$. So, we actually need only investigate the pole structure of $\chi_{RPA}^+$ and $\chi_{RPA}^{zz}$. This is quite easy to do for $\chi^{zz}$, since only one state has a non-vanishing $S^z$ matrix element with the ground state at all $B$. Calling this state $\ket{z}$ and its energy $\epsilon_z$, we find poles of $\chi^{zz}_{RPA}$ at
\begin{align}
\omega_z(\bk) = \epsilon_z \sqrt{1 + \frac{2 J(\bk)}{\epsilon_z}\left|\left\langle 0\middle|\aS^z\middle|z \right\rangle\right|^2}
\label{eq:zeng}
\end{align}
where it should be noted that $\epsilon_z$, $\ket{0}$, $\ket{z}$ and $\aS^z$ all depend on $B$ through the mean field. This result can also be seen to be consistent with perturbation theory to first order in $J_2$, though one must carefully track the dependence of the mean field on $J_2$ in order to obtain all the terms.

\begin{figure}[t]
\includegraphics[scale=0.95]{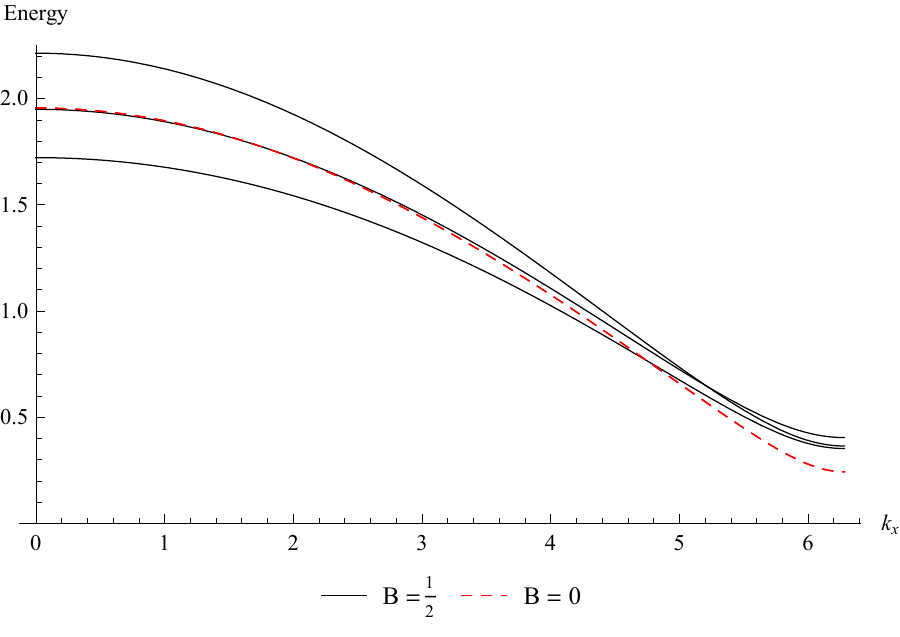}
\caption{Energy of the $t_1$ triplet excitations versus $k_x$ for a few values of $B$, at $J_2 = \frac{\lambda}{17}$. Field and energy are both measured in units of $\lambda$. $B = \frac{1}{2}$ with $\lambda = 22.1$ K corresponds to a physical field of $B_{phys} = 8.23$ T. (See Section~\ref{sec:gfac}).}
\label{fig:disp2}
\end{figure}

The poles of $\chi^+_{RPA}$ are more difficult to extract exactly, due to the larger number of states contributing to $\chi^+_0$. We can cast their location as the roots of an $8$th order polynomial, however, and solve this polynomial numerically as a function of $B$ and $J_2$, producing Figures~\ref{fig:disp} and~\ref{fig:disp2}. Notice that non-zero $J_2$ actually reinforces the linear behavior of the triplet at $\bk = 0$ in small field. This is easily understood by noticing that the mean field is strictly smaller than the applied field, since the antiferromangetic interaction imposes an energy cost to uniform magnetization.

\subsection{The Linear $B$ Regime}
\subsubsection{Computation of the $g$ Factor}
\label{sec:gfac}
In order to characterize the splitting of the magnetic triplet in low field, we can compute an effective $g$ factor in RPA
\begin{align}
g(\bk) = 2 \left.\frac{\partial \epsilon_{+}(B,\bk)}{\partial B}\right|_{B = 0}
\end{align}
where we have anticipated that this splitting may depend on wave vector, and included a factor of $2$ to account for the fact that $B = 2\mu_B B_{phys}$.\cite{opticalabFe-PR,opticalabFe-PRB} Now, if we define the total field felt by the site as
\begin{align}
B_s = B - \beta^z
\end{align}
we then find
\begin{align}
g(\bk) = 2 \left.\frac{\partial \epsilon_{+}(B_s,\bk)}{\partial B_s}\right|_{B_s = 0}\left.\frac{\partial B_s}{\partial B}\right|_{B = 0}
\end{align}
Implicitly differentiating the mean field consistency equation gives
\begin{align}
\left.\frac{\partial B_s}{\partial B}\right|_{B = 0} = \frac{1}{1+ 4 \frac{J(\bzero)}{\lambda}}
\end{align}

For the other derivative, we use first order regular perturbation theory on the polynomial derived from RPA to find
\begin{align}
2 \left.\frac{\partial \epsilon_{+}(B_s,\bk)}{\partial B_s}\right|_{B_s = 0} = 1 + 4\frac{J(\bk)}{\lambda}
\end{align}
so that
\begin{align}
g(\bk) = \frac{\lambda + 4J(\bk)}{\lambda+ 4J(\bzero)}\label{eq:gfactor}
\end{align}

At the zone center, this predicts no modification to the single site $g$ factor of $1$. Curiously, we also find that the $g$ factor at the ordering wave vector decreases to zero as we approach the critical point. This agrees well with the qualitative behavior observed in Figure~\ref{fig:disp2}. Notice also that by $B = 1/2$ and $J_2/\lambda = 1/17$, we are already well outside of the the linear regime at the ordering wave vector. 
\\
\subsubsection{Small Fields}
\label{sec:iso}
In the linear $B$ regime, we actually find that the system responds identically to a static applied field $\bB =B\hat{n}$ in \textit{any} direction. To see this, we show that the change in the ground state magnetization and dynamic susceptibility of the single site is isotropic to first order in $B$. Since these are the only two quantities from the single site problem that enter the calculation of the RPA susceptibility, this is sufficient to show that the response is isotropic at the RPA level. For this section only, $B$ will refer to the magnitude of an arbitrarily directed field, rather than the magnitude of field applied along the $(001)$ direction. So, let $\ket{m}$ denote any exact eigenstate of the single site Hamiltonian in the presence of $\bB$ and write
\begin{align}
\ket{m} = \ket{m_0} + B\ket{m_1}+O(B^2)
\end{align}
For the ground state, we can see
\begin{align}
\ket{0_1} = \frac{1}{\lambda}\sum_{j\neq 0}\ket{j_0} \inoprod{j_0}{\hat{n}\cdot\bS}{0_0}
\label{eq:gsshift}
\end{align}
giving
\begin{align}
\bom = \frac{\inoprod{0}{\bS}{0}}{\inprod{0}{0}^2} =2B\Re\left[\inoprod{0_0}{\bS}{0_1}\right] = \frac{4\bB}{\lambda}+O(B^2)
\end{align}
so that the magnetization of the ground state is indeed isotropic to first order in $\bB$.

As for the single site susceptibility, we will receive two corrections to first order in $B$, one from the first order correction to the energies and one from the first order correction to the states. We will write this as
\begin{align}
\chi_0^{\mu\nu}(\omega,B)  = \chi_0^{\mu\nu}(\omega,0) + B\left(\eta_{E}^{\mu\nu}(\omega) + \eta_{s}^{\mu\nu}(\omega)\right)+O(B^2)
\end{align}
with $\eta_E$ begin the first order correction from a shift in the energies and $\eta_s$ being that from the states. Let us focus first on $\eta_s$. Investigating the relevant product of matrix elements from the spectral representation for $\chi_0$ gives
\begin{align}
\inoprod{0}{S^\mu}{n}\inoprod{n}{S^\nu}{0} =&\inoprod{0_0}{S^\mu}{n_0}\inoprod{n_0}{S^\nu}{0_0} \nonumber\\
&+ B\inoprod{0_1}{S^\mu}{n_0}\inoprod{n_0}{S^\nu}{0_0}\nonumber\\
& + B\inoprod{0_0}{S^\mu}{n_1}\inoprod{n_0}{S^\nu}{0_0}\nonumber\\
& + B\inoprod{0_0}{S^\mu}{n_0}\inoprod{n_1}{S^\nu}{0_0}\nonumber\\
& + B\inoprod{0_0}{S^\mu}{n_0}\inoprod{n_0}{S^\nu}{0_1}+O(B^2)
\end{align}
to first order in $B$. Since some expectation of the form $\inoprod{n_0}{S^\alpha}{0_0}$ appears in each of the summands, we can see that this vanishes to first order in $B$ for all states except those that evolve from members of $t_1$. Furthermore, for $\ket{n_0}\in t_1$, we see that $\ket{n_1}$ is orthogonal to $t_1$. Since $S^\alpha\ket{0_0}$ lies entirely in $t_1$, only the first, second and last summands contribute. Using Equation~\ref{eq:gsshift}, we find
\begin{align}
\eta_s^{\mu\nu} = \frac{4 i}{\lambda}\frac{\omega}{\lambda^2 - \omega^2} n_\alpha \epsilon^{\alpha\mu\nu}
\end{align}
where $\hat{n} = n_\alpha\be_\alpha$ and $\epsilon$ is the Levi-Civita symbol. Indeed, this contribution to the dynamic susceptibility is isotropic.

For $\eta_E$, the matrix elements in the spectral representation are all between zeroth order eigenstates, so we can again restrict our attention to the $t_1$ states. Here we diagonalize the perturbation ($\bB\cdot \bS$) restricted to $t_1$ to obtain the zeroth order eigenstates and the first order energies. Only two states receive corrections to their energies at first order. A straightforward computation then produces
\begin{align}
\eta_E^{\mu\nu} = -\frac{4i\lambda\omega}{(\lambda^2 - \omega^2)^2}n_\alpha\epsilon^{\alpha\mu\nu}
\end{align}
which we can easily see is also isotropic.
\section{Response to Electric Fields}
\label{sec:Eresp}
\subsection{Electric Dipole Excitations of a Single Site}
Since the tetrahedral symmetry of the single site problem does not include inversion, there is actually a single electric dipole allowed transition from the ground state $a_1\rightarrow t_2$. One would imagine that the story for these excitations ought to be similar to that for the magnetic dipole excitations: the peaks in the permeability due to these excitations will persist in the presence of exchange with a shift in position. In this case, it is actually considerably more difficult to make these statements quantitative for a number of reasons. Perhaps most glaringly in contrast to the question of magnetic dipole excitations, we do not know the identity of the operator which couples the single site problem to an electric field ($\bP$). One can demand that such an operator transform as a vector under the point group, i.e. as $t_2$, but this still leaves the magnitude of its five reduced matrix elements undetermined. 

Furthermore, we have no reason to believe that the shift due to the $J_2$ term ought to be dominant over those due to all of the other symmetry allowed exchange terms absent in the $J_2$-$\lambda$ model. Though $J_2$ is thought to dominate the other couplings, it only corrects the energy of the electric triplet at second order. Other symmetry allowed terms,\cite{fescs-prb,fescs-prl} e.g. $(T^y_j\bS_j)\cdot(T^y_i\bS_i)$, give corrections to the energy at first order in their coupling constants that, taken together, might dominate those of $J_2$. Since there has not yet been an experimentally unambiguous observation of the electric triplet excitation, we abandon the question of quantitatively predicting the shift in its energy.

\subsection{Collective Response and Critical Behavior}
\label{sec:critresp}
\subsubsection{Coupling the Critical Theory to Electric Fields}
We can also consider the form of the response to electric field coming entirely from the low-lying magnetic excitations. As we will see, multiple triplon excitations possess the correct symmetry to be produced through coupling to the electric field. To determine the contribution of such processes to the electric field response of our model, we first restrict our considerations to low energy modes near the ordering wave vectors. Either expanding $\Omega_0$ close to the ordering wave vectors and for small frequencies or performing a symmetry analysis\cite{fescs-prb,fescs-prl} produces a Gaussian theory of the form
\begin{align}
\tilde{S}_{eff}\left[\bpsi_{a,\mu}\right] = \frac{1}{\beta }\sum_{a,\mu,\omega_n}\int\frac{\dd^3 \bk}{(2\pi)^3} G_\mu^{-1}(\bk,i\omega_n)\abs{\bpsi_{a,\mu}(\bk,i\omega_n)}^2 
\end{align}
with
\begin{align}
G_\mu^{-1}(\bk,i\omega_n)  =  -(i\omega_n)^2 + \bk V_\mu \bk + r^2 = -(i\omega_n)^2 + \epsilon_\mu^2\label{eq:green}
\end{align}
where $\mu$ labels which ordering wave vector each of the fields came from, and $a$ labels the sublattice. The matrix $V_\mu$ is of the form
\begin{align}
V_x = \begin{pmatrix}
v_1 & 0 & 0 \\
0 & v_2 & 0 \\
0 & 0 & v_2
\end{pmatrix}
\end{align}
with $V_y$ and $V_z$ obtained by permutation. $\bpsi$ is an order parameter for the staggered magnetization, though depending on how we obtained this theory, $\bpsi_{a,\mu}$ may not be precisely the staggered magnetization on the $a$ sublattice at the $\mu$ ordering wave vector. For one thing, we have rescaled the field in order to set the coefficient of the $\omega^2$ term to $1$. Additionally, expectations of $\Phi$ are not precisely those of $S$ (c.f. Equation~\ref{eq:relate}). We expect that such a theory should describe our system correctly on energy scales small compared to the magnetic bandwidth.

Now, rather than investigate the microscopic origins of the coupling of an electric field to this model, we simply investigate which couplings are allowed by symmetry. We expect couplings through a term linear in the applied field, $E^aP^a$, where $P^a$ is some function of the order parameter. This gives rise to a contribution to the $\bk = 0$ electric susceptibility through the standard linear response formalism
\begin{align}
\chi_e^{ab}(\bzero,\omega) = \av{P^a(\bzero,\omega)P^b(\bzero,-\omega)}- \left\langle P^a\right\rangle\left\langle P^b\right\rangle
\end{align}

We do not have couplings before second order in the order parameter, since the order parameter is odd under time reversal while electric fields are even. Beginning at second order without derivatives, we consider what restrictions requiring that a coupling of the form
\begin{align}
C^{a b c}E^a \psi^b\psi^c\equiv E^a P^a_0\label{eq:noderiv}
\end{align}
to transform trivially under the space group places on the tensor $C$, and hence the polarization $P_0^a$. We find that the symmetry allowed coupling is given by
\begin{align}
P_0^x = &c_1 \left(\psi_{A,x}^y\psi_{A,x}^z - \psi_{B,x}^y\psi_{B,x}^z\right)\nonumber\\
& + c_2 \left(\psi_{A,y}^y\psi_{A,y}^z - \psi_{B,y}^y\psi_{B,y}^z+\psi_{A,z}^y\psi_{A,z}^z - \psi_{B,z}^y\psi_{B,z}^z\right)\nonumber\\
& + c_3 \left(\psi_{A,y}^x\psi_{B,y}^y - \psi_{B,y}^x\psi_{A,y}^y+\psi_{A,z}^x\psi_{B,z}^z - \psi_{B,z}^x\psi_{A,z}^z\right)
\end{align}
where $c_1$, $c_2$ and $c_3$ are undetermined by this analysis. The other components are related to the $x$ component by simultaneous permutations of the vector and wave vector indicies. Notice that $P_0$ is odd under interchange of the sublattices, since inversion acts only to interchange the sublattices and gives a sign on the electric field. It can also be shown by expanding the effective action for $\Phi$ given in Equation~\ref{eq:Seff} that a coupling of the electric field to each site through the single site polarization operator ($\bP$) produces a coupling to the critical theory precisely of this form with $c_1 = c_2$ and $c_3 = 0$.

The next lowest order contribution to the electric field response should come from a coupling of the form
\begin{align}
D^{abcd}E^a \psi^b\partial_c\psi^d\equiv E^a P_1^a\label{eq:withderiv}
\end{align}
Due to the larger number of indicies, $P_1^x$ contains many more terms than $P_0^x$ so we omit a detailed discussion of its structure.

\subsubsection{Computation of the Electric Susceptibility}
Neglecting all couplings of higher order in fields and derivatives, we find the electric susceptibility is given by
\begin{align}
\chi_e &=\chi_{e0}+\chi_{e1}\\
\chi_{e0}^{ab} &= \av{P^a_0(\bzero,\omega)P^b_0(\bzero,-\omega)}\\
\chi_{e1}^{ab} & = \av{P^a_1(\bzero,\omega)P^b_1(\bzero,-\omega)}- \left\langle P^a_1\right\rangle\left\langle P^b_1\right\rangle
\end{align}
since $\av{P_0} = 0$ and
\begin{align}
\av{P_0^a(\bzero,\omega)P_1^b(\bzero,-\omega)} = 0
\end{align}
since the internal momentum sum is odd under $\bq\rightarrow - \bq$ while the Green's function is even. Exploiting the particular form of $P_0$, we find that $\chi_{e0}$ is a multiple of the identity and
\begin{align}
\chi_{e0}^{xx}(\bzero,i\omega_n) = \frac{\alpha}{\beta }\sum_{\nu_n}\int \frac{\dd \bq}{(2\pi)^3}G_x(\bq,i\omega_n+i\nu_n)G_x(-\bq,-i\nu_n)
\end{align}
with $\alpha = 2c_1^2 + 4c_2^2 + 4c_3^2$. Performing the Matsubara sum and analytically continuing to real frequencies gives
\begin{align}
\chi_{e0}^{xx}(\bzero,\omega) = \lim\limits_{\delta\rightarrow 0^+}\frac{\alpha}{N}\int \frac{\dd \bq}{(2\pi)^3}\frac{\coth\left(\frac{\beta\epsilon_x(\bq)}{2}\right)}{\epsilon_x(\bq)\left(4\epsilon_x(\bq)^2 - (\omega + i \delta)^2\right)} 
\end{align}
Allowing $\delta \rightarrow 0$ produces
\begin{align}
\text{Im}\left(\chi_{e0}^{xx}(\bzero,\omega)\right)&\propto \text{sgn}(\omega)\int \frac{\dd^3\bq}{(2\pi)^3}\frac{\coth\left(\frac{\beta\epsilon_x(\bq)}{2}\right)}{\epsilon_x(\bq)^2}\delta(\abs{\omega} - 2\epsilon_x)\label{eq:nomom}\\
& \propto \Theta(\abs{\omega} - 2\sqrt{r})\coth(\beta\omega/4)\label{eq:badad}
\end{align}

Now, as for $\chi_{e1}$, we can see on general grounds that
\begin{align}
\chi_{e1}^{ab}(\bzero,i\omega_n) = \frac{\eta^{ab cd}}{\beta }\sum_{\nu_n}\int \frac{\dd\bq}{(2\pi)^3}q_c^2G_{d}(\bq,i\omega_n + i \nu_n)G_d(-\bq,-i\nu_n)
\end{align}
for some fantastically complicated tensor $\eta$. Terms with momentum dependence of the form $q_cq_d$ for $c\neq d$ vanish due to the momentum integration, together with the $\bq\rightarrow - \bq$ symmetry of the Green's function. The restriction that both Green's functions come from the same ordering wave vector comes from demanding that $P_1$ be invariant under the primitive lattice translations, together with the subtraction of $\av{P_1}$. On identical grounds to Equation~\ref{eq:nomom}, we then see
\begin{align}
\text{Im}\left(\chi_{e1}(\bzero,\omega)\right)&\propto \text{sgn}(\omega)\int \frac{\dd^3\bq}{(2\pi)^3}\frac{q_a^2\coth\left(\frac{\beta\epsilon_b(\bq)}{2}\right)}{\epsilon_b(\bq)^2}\delta(\abs{\omega} - 2\epsilon_b)\\
&\propto \Theta(\abs{\omega} - 2\sqrt{r})(\omega^2 - 4r)\coth(\beta\omega/4)\label{eq:goodad}
\end{align}
\begin{figure}[t!]
\includegraphics[scale=0.95]{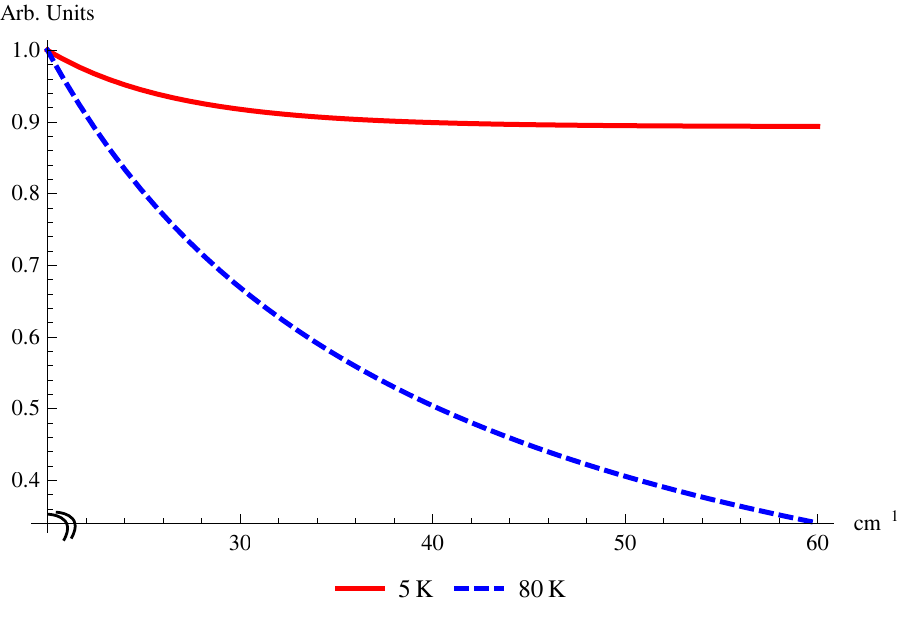}
\caption{Dielectric loss due to Equation~\ref{eq:badad} as a function of wavenumber, in arbitrary units, at 5 K and 80 K. Both functions are normalized so that their value at 20 cm$^{-1}$ is equal to 1.}
\end{figure}
\begin{figure}[t!]
\includegraphics[scale=0.95]{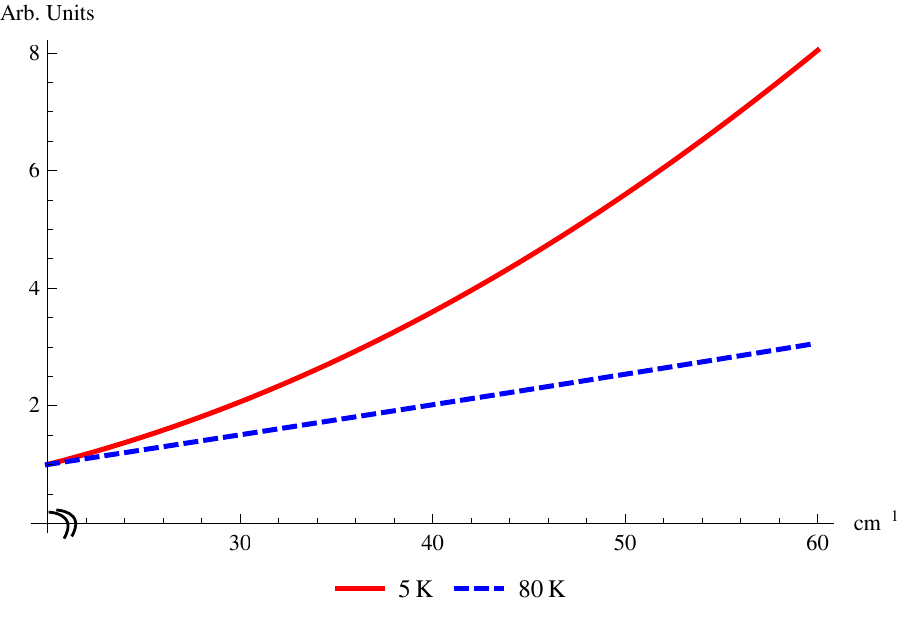}
\caption{Dielectric loss due to Equation~\ref{eq:goodad} as a function of wavenumber, in arbitrary units, at 5 K and 80 K. Both functions are normalized so that their value at 20 cm$^{-1}$ is equal to 1.}
\label{fig:goodad}
\end{figure}
\section{Discussion}
\subsection{Observations of the $a_1\rightarrow t_1$ Excitation}
\subsubsection{Existing Observations}
Two recent THz spectroscopy experiments\cite{ex-cont,ex-gfac} on FeSc$_2$S$_4$ have observed a well defined peak at in the range  of $4.3$meV to $4.5$meV, in broad agreement with the previously estimated magnitude of $\lambda$ and Equation~\ref{eq:old}, which together predict that the $t_1$ excitation at $\bk = 0$ should appear at approximately $3.7$meV.\cite{fescs-prb,fescs-prl} Laurita et al.\cite{ex-gfac} also performed this experiment in field and extracted results in remarkable agreement with the calculations performed here. To briefly recapitulate their story, they were capable of measuring the dynamic susceptibility in the presence of a field with incident light polarized both along the static applied field direction and transverse to it. At zero field, they observed a peak at $\approx4.5$meV in both polarization configurations. As field increased, the peak in the transverse direction split into two peaks in an approximately linear manner while the peak in the longitudinal direction remained unaffected. Fitting a line to the splitting of the peaks in the transverse susceptibility gave them $g \approx 0.92$.

Following the discussion for Section~\ref{sec:magdip}, this is precisely what we would expect for the case of the field along the $(001)$ direction. We saw that the longitudinal susceptibility (i.e. $\chi^{zz}$) received contributions from only one state and, expanding Equation~\ref{eq:zeng} to first order in $B$, the energy of this state is independent of applied field to first order. The transverse susceptibility (i.e. the $x$-$y$ block) received contributions from two states of the lower triplet, whose energies split with a $g$ factor just below the single site value of $g = 1$, due to the small wavevector of the incoming light. A priori, we might be surprised that the analysis with $B$ along the $(001)$ direction fits so well to data taken on a polycrystaline sample, but the results of Section~\ref{sec:iso} tell us that this is exactly what we should expect, provided we are within the linear regime with respect to the static applied field.

We would be remiss if we did not take this time to say a few words about the nature of the $t_1$ excitations and what selection rules are relevant to this situation. It is tempting to draw an analogy between the states of $t_1$ and those of a spin one triplet. Indeed, this analogy motivated the character of the analysis in Section~\ref{sec:magdip}. After all, if we restrict $O(3)$ to $T_d$, the spin one representation becomes the $t_1$ representation. So, at least formally, we can label the members of the triplet by $m = 0$ and $m = \pm 1$. In limited ways, this is even reflected in the response to applied field: the $m = 0$ and $m = \pm 1$ states are the zeroth order eigenenstates with respect to the perturbation $BS^z$ and the first order corrections to the energies are  $0$ and $\pm B/2$, respectively. This is about where the analogy ends, however. The first order corrections to the eigenstates are non-zero (in sharp contrast with a true spin triplet) and there are higher order corrections to the energies of all members of the $t_1$ triplet. 

Furthermore, one might be tempted to attempt to extrapolate selection rules from this analogy, saying that $a_1\rightarrow t_1$ is spin forbidden as a singlet to triplet transition. This would be incorrect, since in terms of the physical spins the transition is between different states of the $S = 2$ manifold, and no total spin change actually occurs. Since the electronic states see a reduced symmetry due to the crystal field and this reduced symmetry is communicated to the spins through spin orbit coupling, neither total spin nor total orbital angular momentum nor total angular momentum are good quantum numbers on energy scales comparable to $\lambda$, and we should not analyze selection rules in terms of them. The proper way to determine the selection rules for transitions between the eigenstates of $\mathcal{H}_i^0$ is through using the Wigner-Eckart Theorem applied to $T_d$.

\subsubsection{Proposal for Future Measurements}
The authors are quite taken with the results of the computation of the $g$ factor (Equation~\ref{eq:gfactor}), and hope that it can be successfully measured soon away from $\bk = \bzero$. This would probably require an inelastic neutron scattering measurement on a single crystal. The benefits of such a measurement would be twofold. First of all, the result is unusual and interesting in itself and it would be valuable to see it confirmed in the material. A successful fit of the $g$ factor to the derived form would argue quite strongly for the predictive power of the $J_2$-$\lambda$ model for this compound. Secondly, and perhaps more importantly, a fit of the $g$ factor to Equation~\ref{eq:gfactor} would provide a measurement of the ratio $J_2/\lambda$, allowing one to estimate the proximity to the critical point directly.
\subsection{Continuum Weight in THz Absorption}
In addition to their observation of a transition matching the description of the $a_1\rightarrow t_1$ excitation, Mittelst\"{a}dt et al.\cite{ex-cont} observed a curious continuum weight at low frequencies in the dielectric loss which was roughly linear at $T = 80$K and superlinear at $T = 5$K. One can attempt to explain this weight in terms of triplon pair production using the analysis in Section~\ref{sec:critresp}, focusing on the $\omega^2\coth(\beta \omega/4)$ term that appears in Equation~\ref{eq:goodad}. At high temperature, $\coth(\beta \omega/4) \approx 4k_B T/\omega$ and we obtain the linear weight. At low temperature, the $\coth$ saturates and we obtain weight that grows as $\omega^2$. Indeed, comparing Figure~\ref{fig:goodad} with the inset in Figure 6 of Mittelst\"{a}dt et al.\cite{ex-cont}, this functional form gives good qualitative agreement with the observed data. Within this picture, the fact that this continuum appears at low frequencies then becomes yet another signal of our proximity to the critical point, since we expect this weight only at frequencies in excess of twice the gap to triplon production at the ordering wave vector, as evidenced by the Heavyside $\Theta$ in Equation~\ref{eq:goodad}.

To be fair, there are a few objections that can be raised to this analysis. One could object that without a temperature dependent prefactor the $\omega^2\coth(\beta\omega/4)$ term fails to reproduce the observed temperature dependence of the absorption. The $\coth$ term decreases with increasing temperature, while the lowest frequency weight observed by Mittelst\"{a}dt et al. increases with increasing temperature. The authors do not find this objection particularly compelling, as one would almost certainly find a temperature dependent prefactor upon a more careful analysis of the temperature dependence. Indeed, if one were to use a finite temperature version of the action in Equation~\ref{eq:Seff} and expand about the ordering wave vector for small frequencies, one would find that the velocities in the Green's function in Equation~\ref{eq:green} depend on temperature. Repeating our analysis that led to Equation~\ref{eq:goodad} but keeping more careful track of constants shows that the velocities appear among the ($\omega$) constant prefactor that we have neglected. This is of course to say nothing of possible temperature dependence of the undetermined constants in the coupling described in Equation~\ref{eq:withderiv}.

Somewhat more concerning is the presence of the two $(const)\times \coth(\beta\omega/4)$ terms appearing in Equations~\ref{eq:badad} and~\ref{eq:goodad}, which do not reproduce the observed $\omega$ dependence. The term in Equation~\ref{eq:goodad} can be argued away consistently by claiming that the gap ($\sqrt{r}$) is very small and we do not expect this term to be easily distinguishable in the presence of the large, dominating $\omega^2$. The contribution from Equation~\ref{eq:badad} cannot, to our knowledge, be so concretely claimed as negligible. One might hope that the presence of small but non-zero $J_1$ might suppress this term by controlling the relative orientations on the $A$ and $B$ sublattices and leading to some cancellation, but this turns out not to be the case. At the Gaussian level, $J_1$ only acts to slightly reorganize our order parameters, shift the gap slightly and move the soft mode slightly away from the ordering wave vector. Pushing through the calculation, one finds only a change to the gap in Equation~\ref{eq:badad}. It seems our only recourse is to argue that the coupling constants ($c_1$, $c_2$ and $c_3$) ought to be small. While this is plausible, the microscopics of this coupling are quite daunting so we do not present a detailed analysis.

\section{Acknowledgments}
This work was supported by the U.S. Department of Energy, Office of Science, Basic Energy Sciences, under Award~\#DE-FG02-08ER46524. The authors would like to thank P. Armitage and N. Laurita for valuable discussions on this topic and L. Mittelst\"{a}dt, P. Lunkenheimer and A. Loidl for a rewarding collaboration on related experimental work.
\bibliography{ref}
\end{document}